\documentclass[aps,pre,twocolumn,superscriptaddress,floatfix]{revtex4}
\usepackage{graphicx}
\usepackage[dvips,unicode,colorlinks,linkcolor=blue,citecolor=blue,urlcolor=blue]{hyperref}
\usepackage{times}

\begin{document}
\title{Mechanical control of heat conductivity in microscopic models of dielectrics}
\author{A. V. Savin}
\email[]{asavin@center.chph.ras.ru}
\affiliation{Semenov Institute of Chemical Physics, Russian Academy of Sciences,
Moscow 119991, Russia}

\author{O. V. Gendelman}
\email[]{ovgend@tx.technion.ac.il}
\affiliation{
Faculty of Mechanical Engineering, Technion -- Israel Institute of Technology,
Haifa 32000, Israel}

\date{\today}

\begin{abstract}
We discuss a possibility to control a heat conductivity in simple one-dimensional models
of dielectrics
by means of external mechanical loads. To illustrate such possibilities we consider
first a well-studied chain with degenerate double-well potential of the interparticle interaction.
Contrary to previous studies, we consider varying length of the chain with fixed number of
particles. Number of possible energetically degenerate ground states strongly depends on the
overall length of the chain, or, in other terms, on average length of the link between neighboring
particles. These degenerate states correspond to mechanical equilibrium, therefore one can say
that the transition between them mimics to some extent a process of plastic deformation. We
demonstrate that such modification of the chain length can lead to quite profound
(almost five-fold) reduction of the heat conduction coefficient. Even more profound effect is
revealed for a model with single-well non-convex potential. It is demonstrated that in certain
range of constant external forcing this model becomes "effectively"\ double-well, and has a
multitude of possible states of equilibrium for the same value of the external load. Thus,
the heat conduction coefficient can be reduced by two orders of magnitude.
We suggest a mechanical model of a chain with periodic double-well potential,
which allows control over the heat conduction. The models considered
may be useful for description of heat transport in biological macromolecules and for control
of the heat transport in microsystems.

\end{abstract}

\maketitle

\section{Introduction}
Heat conduction in low-dimensional systems has attracted a lot of attention and has been a
subject of intensive studies \cite{LLP03}. The main objective here is to derive from first
principles (on the atomic-molecular level) the Fourier law -- proportionality of the heat flux
to the temperature gradient $J=-\kappa \nabla T$, where $\kappa$ is the heat conduction coefficient.
To date, there exists quite extensive body of works devoted to the numerical modeling of the heat
transfer in the one-dimensional chains. Anomalous characteristics of this process are well-known
since celebrated work of Fermi, Pasta and Ulam  \cite{FPU}. In integrable systems (harmonic chain,
Toda lattice, the chain of rigid disks) the heat flux $J$ does not depend at all on the chain
length $L$, therefore, the thermal conductivity formally diverges. The underlying reason for that
is that the energy is transferred by non-interacting quasiparticles and therefore one cannot
expect any diffusion effects. Non-integrability of the system is a necessary but not  sufficient
condition to obtain the  convergent heat conduction coefficient. Well-known examples are
Fermi-Pasta-Ulam (FPU) chain\cite{LRP,LLP1,LLP2}, disordered harmonic chain \cite{RG,CL,D},
diatomic 1D gas of colliding particles \cite{D1,STZ02,GNY} and the diatomic Toda lattice \cite{H}.
In these non-integrable systems also have divergent heat conduction coefficient; the latter
diverges as a power function of length: $\kappa\sim L^\alpha$, for $L\rightarrow\infty$.
The exponent lies in  the interval $0<\alpha <1$.

On the other side, the 1D lattice with on-site potential can have finite conductivity.
The simulations had demonstrated the convergence of the heat conduction coefficient for
Frenkel-Kontorova chain \cite{HLZ98,SG03}, the chain with hyperbolic sine on-site potential \cite{TBSZ},
the chain with $\phi^4$ on-site potential \cite{HLZ00,AK00} and for the chain of hard disks of
non-zero size with substrate potential \cite{GS04}.
The essential feature of all these models is existence of an external potential modeling the
interaction with the surrounding system. These systems are not translationally invariant, and,
consequently, the total momentum is not conserved. In paper \cite{H} it has been suggested that
the presence of an external potential plays a key role to ensure the convergence of the heat
conductivity.  This hypothesis has been disproved in works \cite{giardina,savin2},
where it was shown that the isolated chain of rotators (a chain with a periodic potential
interstitial interaction) has a finite thermal conductivity.
In recent studies \cite{ZZWZ12,CZWZ12}
it was demonstrated that the heat conduction is convergent even in a chain with Morse potential
of the nearest-neighbor interaction. It seems that this finding is intimately related to the fact
that the "extension" branch of the potential function is finite -- in other terms, a dissociation
of the neighboring particles is possible \cite{SK13}. Thus, the heat flux is scattered on these
fluctuating large gaps between the fragments of the chain.

In the systems mentioned above, the strong non-homogeneities, which critically effect the heat
transfer, are caused by the thermal fluctuations. In current work, we would like to explore
somewhat different idea -- to design the interaction potential and external conditions in a way
that the inhomogeneities will appear in controllable manner and with desired density. Thus,
it might be possible to control the heat conduction coefficient in wide range by simple variation
of the external conditions -- for instance, by stretching the chain.
In order to accomplish this goal, we study a chain with a double-well (DW) potential of the
nearest-neighbor interactions. We also study certain modification of the DW model, which has
only one minimum but can acquire the double-well structure under action of external force.
These systems are studied both under nonequilibrium conditions using Langevin thermostats,
and within the framework of equilibrium molecular dynamics using Green-Kubo formula \cite{GK}.

In the first time the thermal conductivity of the chain with double-well potential was considered
in \cite{giardina}, with the of help nonequilibrium molecular-dynamics simulation with Nose-Hoover
thermostats. It was shown that the use of Nose-Hoover thermostat for the non-equilibrium problems
can be misleading \cite{FHLZ, LLM}. More accurate modeling of the heat transfer using c
Langevin thermostat was presented in \cite{Roy}.

It seems that neither of these previous works considered the relationship between the variations
of the chain length and the thermal conductivity.
However, an important feature of the chain with the DW potential is the existence of a
large number of possible ground states of the chain. So, the chain with $N$ particles and periodic
boundary conditions has $2^{N-1}$ possible ground states with the same energy; the only possible
difference between these states is the overall equilibrium length of the chain. We show that the
thermal conductivity of the chain depends essentially on its ground state, governed by the length.
In each particular simulation, the overall length of the chain is fixed, and all ground states
corresponding to this particular value of the length are considered to be equivalent.
So, we are going to show that a variation of the chain length brings about significant change
of its thermal conductivity. We also show that the same (and even much stronger) effect can be
achieved for special design of a single-well nearest-neighbor potential; in this case, the
multiplicity of the ground states is achieved by application of a uniform external stretching.
Besides, we demonstrate that also the phenomena related to a non-equilibrium heat conduction,
like relaxation modes of thermal perturbations, are strongly affected by the number
of competing ground states

It should be mentioned that the heat conduction coefficient of some models considered in this paper
is believed to be divergent in the thermodynamic limit. This point is not that significant here,
since we discuss the effect of length, stretching and number of the ground states in a chain with
fixed number of particles. Therefore, the heat conduction coefficient is well-defined
in all considered cases.

\section{Description of the model}
Let us consider a general chain with $N$ particles. In a dimensionless form the Hamiltonian
of the chain can be written as
\begin{equation}
H=\sum_{n=1}^N\frac12\dot{u}_n^2+\sum_{n=1}^{N-1}V(u_{n+1}-u_{n}),
\label{f1}
\end{equation}
where $N$ is the total number of particles in the chain, $u_n$ is a coordinate of the $n$-th particle,
dot denotes differentiation with respect to dimensionless time $t$, $V(\rho_n)$ is the nearest-neighbor
interaction potential, $\rho_n$ is the length of the $n$-th link between the neighboring particles.
The coordinate of the particle $u_n$ can not only describe the position of the particles with respect
to the the chain axis; it may also correspond to the rotation angle of the $n$-th
monomer around the rotation axis. In this sort of models $\rho_n$ will denote the relative angle
between the $(n + 1)$-st and the $n$-th monomer.

\begin{figure}[t,b]
\includegraphics[angle=0, width=1\linewidth]{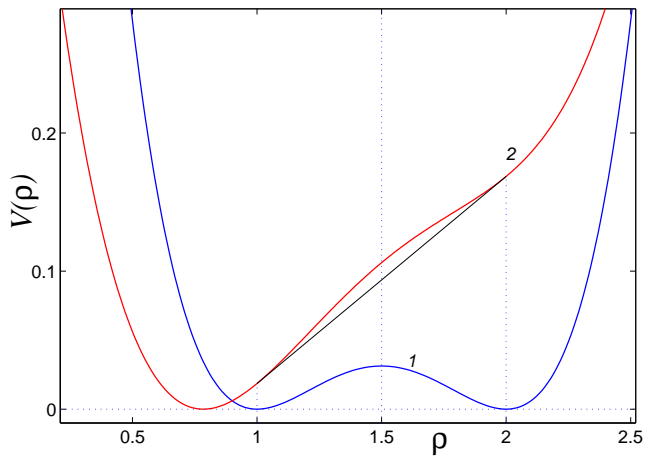}
\caption{ Sketch of double-well potential (\ref{f2}) (curve 1) and
single-well potential (\ref{f6}). The straight line connecting the
the points $(1,V(1))$ and $(2,V(2))$ of graph single-well potential sets its convex hull.
}
\label{fg01}
\end{figure}

We choose the double-well (DW) potential of the interparticle interaction in the following form:
\begin{equation}
V(\rho)=\epsilon (\rho-1)^2(\rho-2)^2,
\label{f2}
\end{equation}
where we choose $\epsilon=1/2$; it leads to $V''(1)=V''(2)=1$. The shape of the potential is
presented in Fig.~\ref{fg01}. The height of the barrier between the minima of the potential
$E_0=V(1.5)=1/2^5=0.0313$.

To simulate the heat transfer in the chain, we use the stochastic Langevin thermostat.
The chain has in general $N_++N+N_-$ particles.
We connect $N_+$ particles from one side of the chain to a "hot"\ Langevin thermostat
with temperature $T_+$, and $N_-$ particles from the other side -- to the Langevin thermostat
with temperature $T_-$. The corresponding system of equations of motion of the chain
can be written as:
\begin{eqnarray}
\ddot{u}_n&=&-\partial H/\partial u_n -\gamma\dot{u}_n+\xi_n^+,~~
n\le N_+,\nonumber\\
\ddot{u}_n&=&-\partial H/\partial u_n,~~n=N_++1,...,N_++N,\label{f3}\\
\ddot{u}_n&=&-\partial H/\partial u_n -\gamma\dot{u}_n+\xi_n^-,~~
n> N_++N, \nonumber
\end{eqnarray}
where $\gamma = 0.1$ is a relaxation coefficient, $\xi_n^\pm$  models a white Gaussian noise
normalized by the conditions
$\langle\xi_n^\pm(t)\rangle=0$,
$\langle\xi_n^+(t_1)\xi^-_k(t_2)\rangle=0$,
$\langle\xi_n^\pm(t_1)\xi_k^\pm(t_2)\rangle=2\gamma T_\pm\delta_{nk}\delta(t_2-t_1)$.

The system of equations of motion (\ref{f3}) was integrated numerically. Considered a chain with fixed
ends: $u_1\equiv 0$, $u_N\equiv (N-1)a$. Used on-initial condition
$$
\{u_n(0)=(n-1)a,~\dot{u}_n(0)=0\}_{n=1}^N,
$$
where $1 < a < 2$ -- average value of the chain link length.
After an initial transient, thermal equilibrium with the thermostats was established and a
stationary heat flux along the chain appeared. A local temperature is numerically defined as
$T_n=\langle\dot{u}_n^2\rangle_t$
and the local heat flux -- as
$J_n=\bar{a}\langle j_n\rangle_t$, where $j_n=-\dot{u}_nV'(u_{n}-u_{n-1})$.
In numerical simulations we used the following values of temperature $T_\pm=(1\pm 0.1)T$
($T=0.01$, 0.03, 0.1), the relaxation coefficient $\gamma = 0.1$, the number of
units $N_\pm=40$, $N=20$, 40, 80, ..., 10240.

\begin{figure}[tb]
\includegraphics[angle=0, width=1\linewidth]{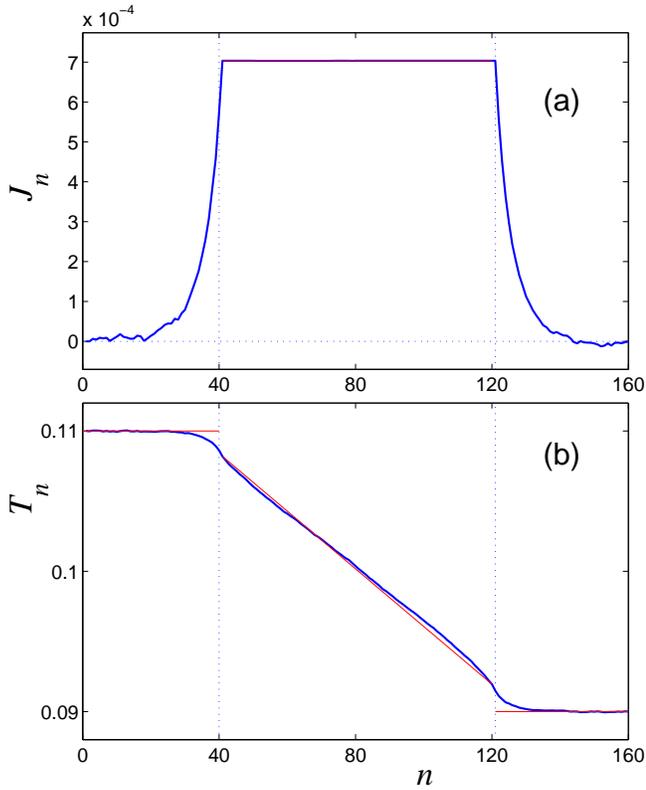}
\caption{
Profile of (a) local heat flux $J_n$ and (b) local temperature $T_n$
in the chain with DW potential (\ref{f2}) for $N_\pm=40$, $N=80$, $T_+=0.11$, $T_-=0.09$.
Linear approximation of the temperature distribution (thin red line) is used for evaluation
of the heat conduction coefficient $\kappa(N)$.
}
\label{fg02}
\end{figure}

This method of thermalization overcomes the problem of the thermal boundary resistance.
The distribution of the heat flow  $J_n$ and the temperature in the chain $T_n$ (Fig.~\ref{fg02})
clearly demonstrates that inside the "internal" fragment of the chain $N_+<n\le N_++N$ we observe
a heat flux independent on the chain cite, as one would expect in an energy-conserving system.
Temperature profile also is almost linear. Thus, from this such simulation one can unambiguously
define the heat conduction coefficient of the internal chain fragment:
\begin{equation}
\kappa(N)=J(N-1)/(T_{N_++1}-T_{N_++N}).
\label{f4}
\end{equation}

In thermodynamical limit, one can say that the system obeys the Fourier law,
if there exists a finite limit
$$
\bar\kappa=\lim_{N\rightarrow\infty}\kappa(N).
$$
If such limit does exist, one can say that the chain has normal (finite) or convergent heat
conduction. In the opposite case, an anomalous heat conduction is observed.

Alternative way for evaluation of the heat conduction coefficient is based on linear response
theory, which leads, in particular, to famous Green-Kubo expression \cite{GK}:
\begin{equation}
\kappa_c=\lim_{t\rightarrow\infty}\lim_{N\rightarrow\infty}\frac{1}{NT^2}\int_0^tc(\tau)d\tau,
\label{f5}
\end{equation}
where the autocorrelation function of the heat flux in the chain is defined as
$c(t)=\langle J_s(\tau)J_s(\tau-t)\rangle_\tau$.
Here a total heat flux in the chain is defined as $J_s(t)=\sum_n j_n(t)$.

In order to compute the self-correlation function $c(t)$ we simulate a cyclic chain consisting of
$N=10^4$ particles and couple all these particles to the Langevin thermostat with temperature
$T$. After initial thermalization, the chain has been detached from the thermostat and
Hamiltonian dynamics has been simulated. In order to improve the accuracy, the self-correlation
function has been computed by averaging over $10^4$ realizations with independent initial
conditions, corresponding to the same initial temperature $T$.

The heat conduction turns out to be convergent if the self-correlation function $c(\tau )$
decreases fast enough as $\tau\rightarrow\infty$. Namely, if the integral in expression (\ref{f5})
converges then the heat conduction may be treated as normal.
\begin{figure}[tb]
\includegraphics[angle=0, width=1\linewidth]{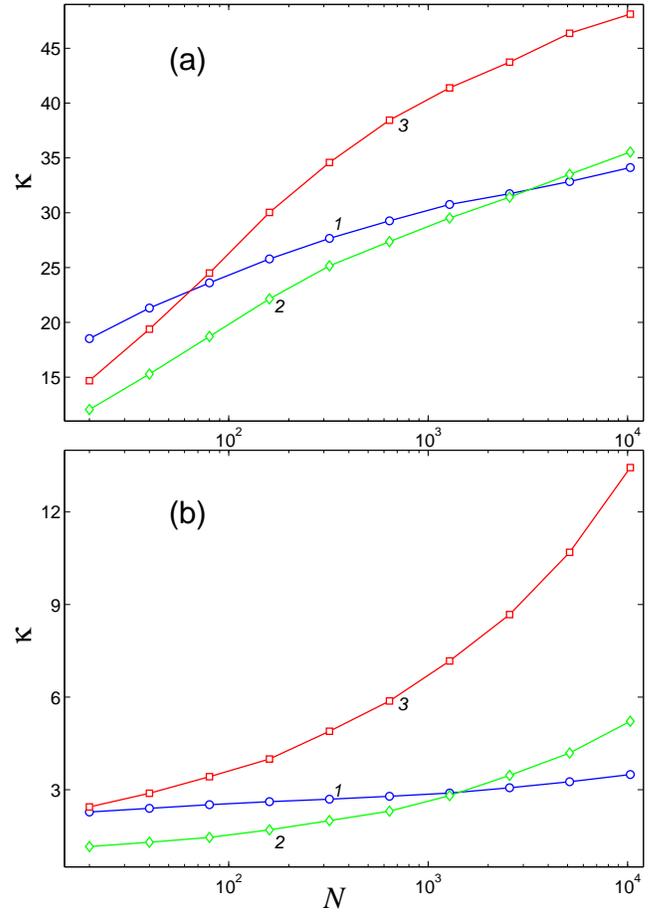}
\caption{
Dependence of the heat conduction coefficient $\kappa$ on the size of the internal chain fragment
$N$ for the chain with the DW potential (\ref{f2}) for the average interatomic distance (a) $a=1$
and (b) $a=1.5$ for temperatures $T=0.01$, 0.03, 0.1 (curves 1, 2, 3 respectively).
}
\label{fg03}
\end{figure}

\section{Chain with symmetric DW potential}

\begin{figure}[tb]
\includegraphics[angle=0, width=1\linewidth]{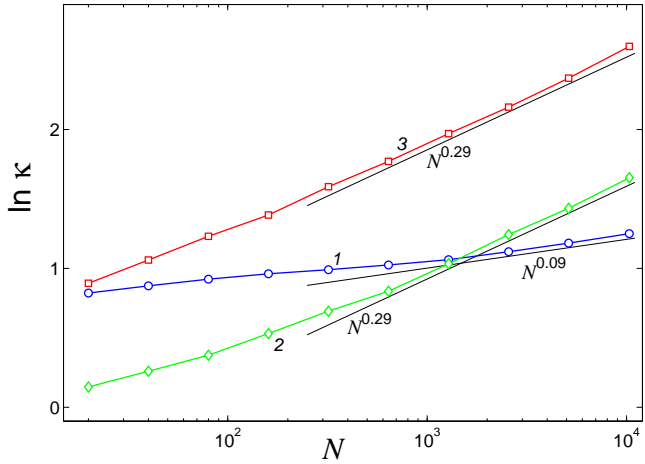}
\caption{
Dependence of the heat conduction coefficient $\kappa$ on the size of the internal chain fragment
$N$ for the chain with the DW potential (\ref{f2}) for the average interatomic distance
$a=1.5$ for temperatures $T=0.01$, 0.03, 0.1 (curves 1, 2, 3 respectively).
The graphs are presented in double logarithmic scale and straight lines correspond
to fitting $N^{\alpha}$, $\alpha=0.09, 0.29, 0.29$
for curves 1, 2, 3 respectively.
}
\label{fg04}
\end{figure}

Dependence of the heat conduction coefficient $\kappa$ on the length of the chain fragment between
the Langevin thermostats $N$ for the chain with symmetric DW potential (\ref{f2}) is presented
in Fig.~\ref{fg03}.
As one can see from this figure, both for $a=1$ and $a=1.5$ the heat conduction coefficient gown
monotonously with $N$ and demonstrates no trend for convergence. For $a=1.5$ the heat conduction
coefficient grows like $N^\alpha$, with $\alpha=0.09$ for the temperature $T=0.01$
and $\alpha=0.29$ for $T=0.03$, 0.1, as it is demonstrated in Fig. \ref{fg04}.

\begin{figure}[tb]
\includegraphics[angle=0, width=1\linewidth]{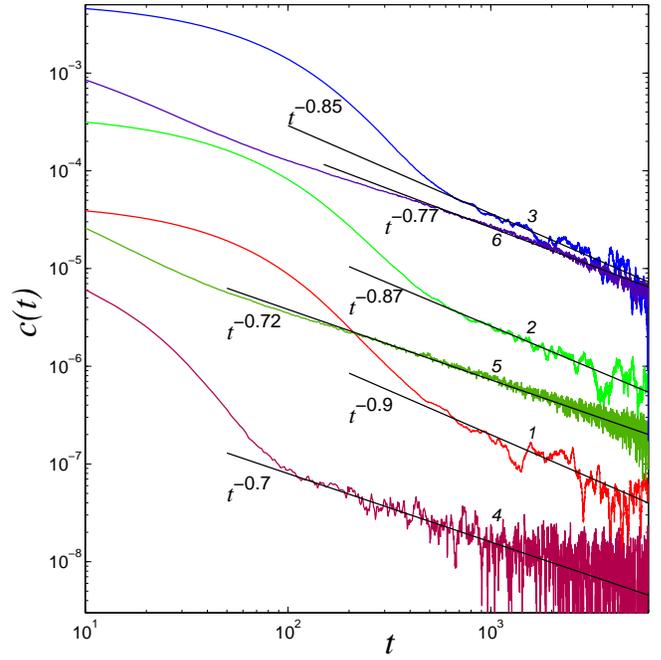}
\caption{
Power-law decrease of the autocorrelation function $c(t)$ for the chain with symmetric DW
potential (\ref{f2}) for average link length $a=1$ and temperatures $T=0.01$, 0.03, 0.1
(curves 1, 2, 3 respectively) and $a=1.5$, $T=0.01$, 0.03, 0.1 (curves 4, 5, 6 respectively).
The graphs are presented in double logarithmic scale and straight lines correspond
to fitting $t^{-\beta}$, $\beta=0.9, 0.87, 0.85, 0.7, 0.72, 0.77$
for curves 1, 2,..., 6 respectively.
}
\label{fg05}
\end{figure}

Numeric analysis of the heat flux autocorrelation function $c(t)$ as $t\rightarrow\infty$ supports
the conclusion on divergent heat conductivity in the system. For all values of the average link
length $1\le a\le 2$ the function $c(t)$ decreases as $t\rightarrow\infty$ according
to the power law $t^{-\beta}$ with exponent $\beta<1$, as it is demonstrated in Fig. \ref{fg05}.
According to Green-Kubo formula (\ref{f5}) one arrives to the same conclusion on divergence
of the heat conduction coefficient $\kappa(N)$ as $N\rightarrow\infty$. Two approaches used here
are independent and point in the same direction -- one obtains clear indication that the heat
conductivity in the chain with DW potential diverges.

\begin{figure}[tb]
\includegraphics[angle=0, width=1\linewidth]{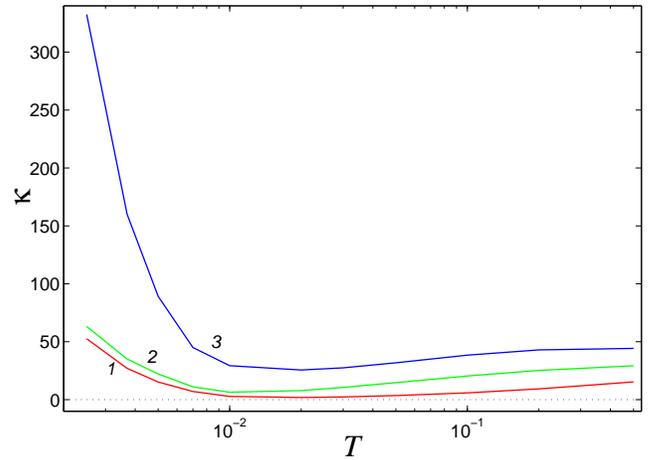}
\caption{
Dependence of the heat conduction coefficient $\kappa$ on temperature $T$ of the chain containing
in general $720$ particles ($N=640$, $N_\pm=40$) with the DW potential (\ref{f2})
for $a=1.5$, 1.1, 1.0 (curves 1, 2, 3 respectively).
}
\label{fg06}
\end{figure}

So, as it was already mentioned in the Introduction, it is reasonable to explore and compare the
heat conduction coefficient $\kappa(N)$ for some fixed chain length. To be specific, we choose
$N=640$ and consider the temperature dependence of the heat conductivity for fixed number of
particles and varying average link length. The results are presented in Fig.~\ref{fg06}.
For all presented values of the link length the heat conductivity in the case of low temperatures
$T<0.01$ sharply decreases as the temperature grows. For large temperatures the heat conductivity
weakly increases with the temperature. In all cases the minimum is achieved in the temperature
interval $0.01 \div 0.04$, which is close to the height of the potential barrier.
The lowest values of the heat conduction coefficient are obtained for $a=1.5$.
This result is quite expectable, since namely for this value of the average link length the
number of topologically different degenerate ground states achieves a maximum.

From Fig.~\ref{fg06} one can see that if the average link length $a$ varies from 1 to 1.5,
the heat conduction coefficient decreases monotonously.
In order to estimate the efficiency of the heat conductivity reduction, we define the reduction
coefficient $\mu_a=\kappa(a,N,T)/\kappa(1,N,T)$, where $\kappa(a,N,T)$ denotes the heat conduction
coefficient for the average link length $a$, number of particles $N$ and temperature $T$.
We note that by symmetry considerations it is enough to consider only the interval $1\le a\le 1.5$.

\begin{figure}[tb]
\includegraphics[angle=0, width=1\linewidth]{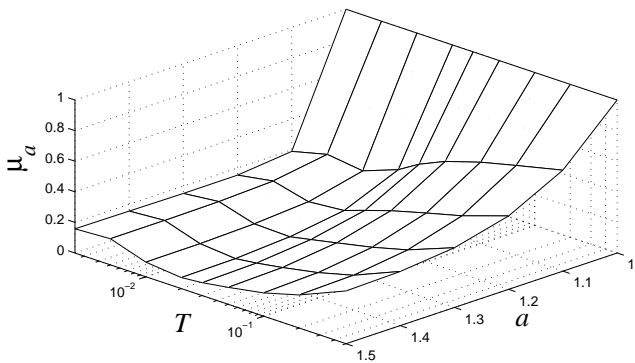}
\caption{
Temperature dependence of the reduction coefficient
$\mu_a=\kappa(a)/\kappa(1)$ on the average link length $a$.
}
\label{fg07}
\end{figure}

A dependence of $\mu_a$ on the temperature and the average link length is presented in Fig.~\ref{fg07}.
The heat conductivity decreases with increase of $a$ for all studied values of the temperature.
Maximal efficiency of the reduction is achieved for the temperature $T=0.02$, which is close
to the height of the potential barrier $E_0=0.0313$. As one could expect, maximal reduction of the
heat conductivity occurs at $a=1.5$.

\begin{figure}[tb]
\includegraphics[angle=0, width=1\linewidth]{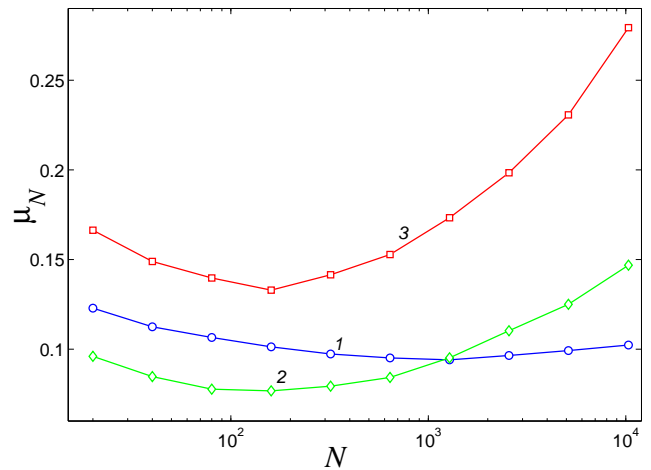}
\caption{
Dependence of the reduction coefficient
$\mu_N=\kappa(1.5)/\kappa(1)$ on the chain length $N$ for temperature values
$T=0.01, 0.03, 0.1$ (curves 1, 2, 3 respectively).
}
\label{fg08}
\end{figure}

However, other aspect of the reduction phenomenon is somewhat unexpected. In Fig.~\ref{fg08}
we present the dependence of the reduction efficiency for different temperatures
$\mu_N=\kappa(1.5,N,T)/\kappa(1,N,T)$ as a function of the particle number $N$. It is somewhat
surprising to see that this dependence is not monotonous. The chain length for which the most
efficient reduction of the heat conductivity is observed, strongly depends on the temperature.
For instance, for $T=0.01$ the most efficient reduction is observed when $N_0=1280$,
and for higher temperatures $T=0.03, 0.1$ -- when $N_0=160$.

It is easy to explain why for relatively short chains the reduction efficiency is higher
when the chain gets longer. We believe that the reduction of the heat conductivity occurs due
to increase of a number of the degenerate ground states. Quite obviously, this effect becomes more
profound as the number of particles grows. The decrease of the reduction efficiency for relatively
high $N$ is more difficult to explain. Qualitatively, one can speculate that, since the heat
conduction coefficient in the chain with DW potential diverges, for very large $N$ the heat
transfer is governed by long-wavelength weakly interacting phonons. Such waves are less sensitive
to the details of the chain structure and "feel"\  only average density of the particles.

It is  also instructive to compare the process of non-stationary heat conduction in the DW model
with different average link length. For this sake, following a methodology developed in
\cite{GSNS1, GSNS2}, we simulate a relaxation of various spatial modes of a thermal perturbation
in the cyclic DW chain with $N$ particles. Thus, the initial temperature distribution is defined
according to the following relationship:
\begin{equation}
T_n=T_0+A\cos[2\pi(n-1)/Z], \label{fa1}
\end{equation}
where $T_0$ is the average temperature, $A$ is the amplitude of the perturbation,
and $Z$ is the length of the mode (number of particles). The overall number
of particles $N$ has to be multiple of $Z$ in order to ensure the periodic
boundary conditions.

In order to realize this initial thermal perturbation, each particle in the chain is first
attached to a separate Langevin thermostat. In other terms, we integrate numerically the following
system of equations:
\begin{equation}
\ddot{u}_n=-\partial H/\partial u_n -\gamma\dot{u}_n+\xi_n, \label{fa2}
\end{equation}
where $n=1,2,...,N$, $\gamma=0.1$, à and the action of the thermostat is simulated as white
Gaussian noise normalized according to conditions
$$
\langle \xi_n(t_1)\xi_k(t_2)\rangle=2\gamma T_n\delta_{kn}\delta(t_2-t_1).
$$
After the initial heating in accordance with (\ref{fa1}), the Langevin
thermostat is removed and relaxation of the system to a stationary
temperature profile is studied. The results were averaged
over $10^6$ realizations of the initial profile $\{T_n\}_{n=1}^N$
in order to reduce the effect of fluctuations.
\begin{figure}[tb]
\includegraphics[angle=0, width=1\linewidth]{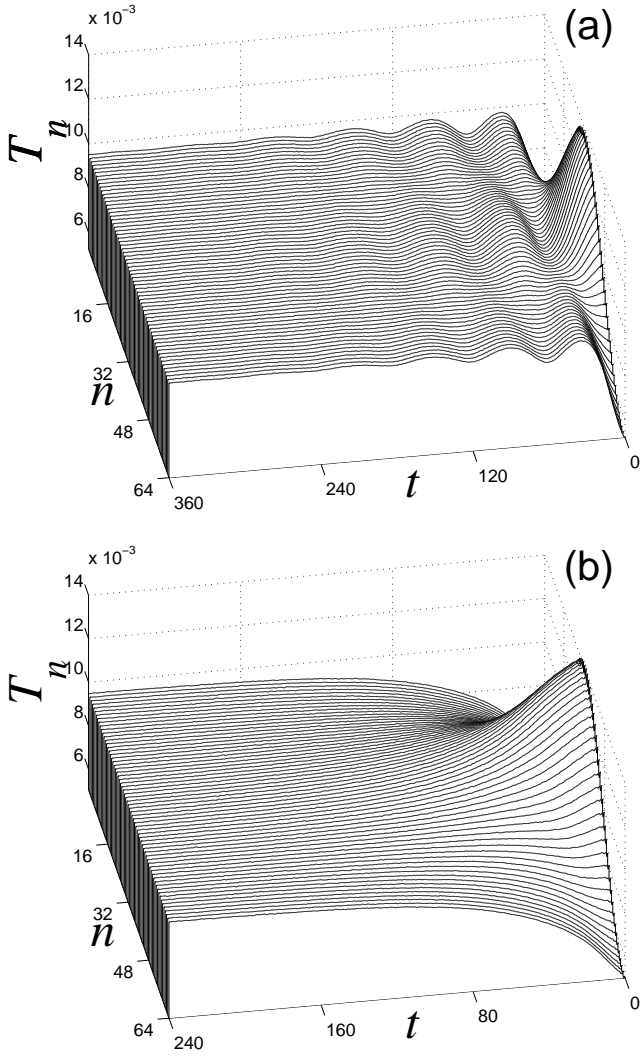}
\caption{
Relaxation of initial periodic thermal profile in the
periodic chain with potential (\ref{f2}) $N=1024$, $T_0 =0.01$, $A = 0.05$,
$Z=64$ for (a) $a=1$ and (b) $a=1.5$.
}
\label{fg09}
\end{figure}
\begin{figure}[tb]
\includegraphics[angle=0, width=1\linewidth]{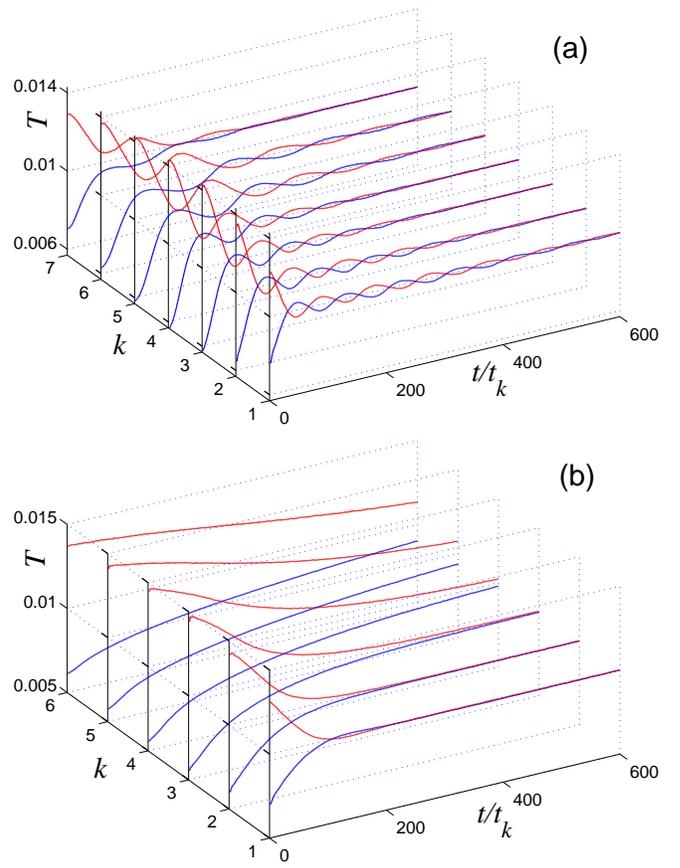}
\caption{
Evolution of the relaxation profile in
the periodic chain with double-well potential (\ref{f2})
($N=2048$, $T_0=0.01$, $A=0.05$) for (a) $a=1$
($Z =2^{3+k}$, $k=1,2,...,7$, $t_k=0.15$, 0.3, 0.6, 1.2, 2.0, 4.0, 10.0)
and (b) $a=1.5$
($Z =2^{3+k}$, $t_k=0.1\cdot 2^{k-1}$, $k=1,2,...,6$).
Time dependence of the mode maximum $T(1+Z/2)$ [red (gray) lines]
and minimum T(1) [blue (black) lines] are depicted.
}
\label{fg10}
\end{figure}

Samples of the non-stationary simulations are presented in Figs.~\ref{fg09} and \ref{fg10}.
In both figures, we compare the relaxation of similar thermal perturbation profiles with similar
average temperature, perturbation amplitude and spatial wavelength. The only difference is the
average link length. We compare the two extreme cases $a=1$ and $a=1.5$. In Fig.~\ref{fg10}
one can see that the same thermal perturbation for $a=1$ decays in oscillatory manner,
whereas for $a=1.5$ the decay is smooth. Thus, one can suggest that for the case $a=1$ one
should take into account phenomena related to hyperbolic corrections to Fourier law
(like possibility of the second sound); in the same time, for the case $a=1.5$ one should expect
primarily diffusive behavior. This conclusion is further confirmed by simulation results presented
in Fig.~\ref{fg09}. There we can see that for $a=1$ the oscillatory (i.e. hyperbolic) behavior
is observed for very broad diapason of the wavelengths. For all these modes, the thermal
perturbations for $a=1.5$ decay in primarily diffusive manner. The reason for this difference,
again, is a "perfect" structure of the ground state for $a=1$ and large number
of the degenerate ground states for $a=1.5$.

\section{A model with non-convex single-well potential}

The results presented in the previous Section suggest that the increase of the number of
the ground states peculiar for the DW chain leads to suppression of the heat conduction.
One should expect even stronger suppression effect, when the chain is modified in a way that
the potential changes from single-well to double-well. Possibility of such modification
in particular chain model under action of an external force, and the consequences for the heat
conductivity are discussed here.
First of all, we suggest a model of the chain with a non-convex single well potential
of the nearest-neighbor interaction:
\begin{equation}
V(\rho)=\epsilon (\rho-1)^2(\rho-2)^2+\beta\rho+c,
\label{f6}
\end{equation}
where $\epsilon=1/2$, $\beta=0.15$, à and (physically insignificant) constant
$c$ may be found from a condition $\min V(\rho)=0$. Under these values of parameters potential
function (\ref{f6}) has a single minimum at $\rho_0=0.89$ (Fig.~\ref{fg01}).

Let us consider the case when the chain with potential (\ref{f6}) is elongated by external force
$F$ applied to one of its ends, whereas the other end remains fixed. It is easy to see that
such external forcing is equivalent to a modification of the interaction potential (\ref{f6})
by addition of negative linear term. This modified potential will have the following form:
\begin{equation}
V^{*}(\rho)=\epsilon (\rho-1)^2(\rho-2)^2+\beta\rho-F\rho+c.
\label{f6new}
\end{equation}
Thus, the application of the external force can modify the qualitative shape of the potential
function. Namely, it is easy to demonstrate that for
$F<F_{min}=\beta-\epsilon/3\sqrt{3}$ and for $F>F_{max}=\beta+\epsilon/3\sqrt{3}$
the potential $V^{*}(\rho)$ remains single-well. However for $F_{min}<F<F_{max}$ the effective
potential becomes double-well and thus one can expect significant reduction of the heat
conductivity as a result of application of the external force.

In our numeric experiments we control the average link length $a$ rather than the value of the
external force. However, it is easy to translate between these quantities. Namely, potential
function (\ref{f6new}) will have two wells if the following condition holds:
$a_{min}<a<a_{max}$,
where
$a_{min}=3/2-1/\sqrt{3}\approx 0.9226$ and $a_{max}=3/2+1/\sqrt{3}\approx 2.0774$.

Creation of the effective double-well potential can be also explained from observation of the
single-well potential function (\ref{f6}) depicted in Fig.~\ref{fg01}. One can see that this
function is not convex (the second derivative is negative for $1\le\rho\le 2$). Homogeneous
extension of such chain, when all the links have the same length $\rho_n\equiv a>\rho_0$, is
energetically favorable only when the average link length belongs to the interval where the
potential function has positive second derivative: $a\le\rho_1=1$ and $a\ge\rho_2=2$.
If $\rho_1<a<\rho_2$, it is more favorable to have part of the links with the length
$\rho_n\equiv\rho_1=1$, and the rest of the links -- with the length $\rho_n\equiv\rho_2=2$.
In this case, the dependence of the average potential energy per particle on the average link
length will follow the straight line connecting the points
$(\rho_1,V(\rho_1))$ and $(\rho_2,V(\rho_2))$,
as it is demonstrated in Fig.~\ref{fg01} \cite{Savin_pnas,smko13}. Thus, as the average link
length increases, the numbers of "short" and "long" links vary accordingly, thus giving rise
to large number of possible realizations and large variations of the heat conduction coefficient.

Heat transfer in this system has been simulated with the help of Langevin thermostats (\ref{f3})
under fixed-ends boundary conditions: $u_1(t)\equiv 0$, $u_{N_++N+N_-}=(N_++N+N_--1)a$,
where again $a\ge\rho_0$ is the average length of the link.
\begin{figure}[tb]
\includegraphics[angle=0, width=1\linewidth]{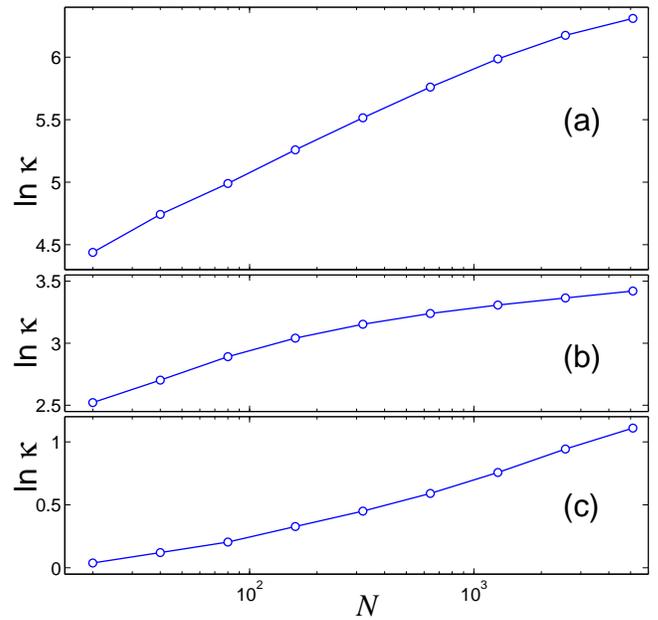}
\caption{
Dependence of heat conduction coefficient $\kappa$ on the chain length $N$
for the chain with single-well nearest-neighbor potential (\ref{f6}) for different
values of the average link length:
(a) $a=\rho_0$, (b) $a=2.0$, (c) $a=1.5$. In all cases the temperature $T=0.01$.
}
\label{fg11}
\end{figure}

Numeric simulation of the heat conduction demonstrated that for all studied values
of the average link length $a\ge\rho_0$ the heat conduction coefficient diverges:
$\kappa(N)\nearrow\infty$ as $N\nearrow\infty$ -- see. Fig.~\ref{fg11}.
Equilibrium simulations based on Green-Kubo formula (\ref{f5}) bring about the same conclusion.
\begin{figure}[tb]
\includegraphics[angle=0, width=1\linewidth]{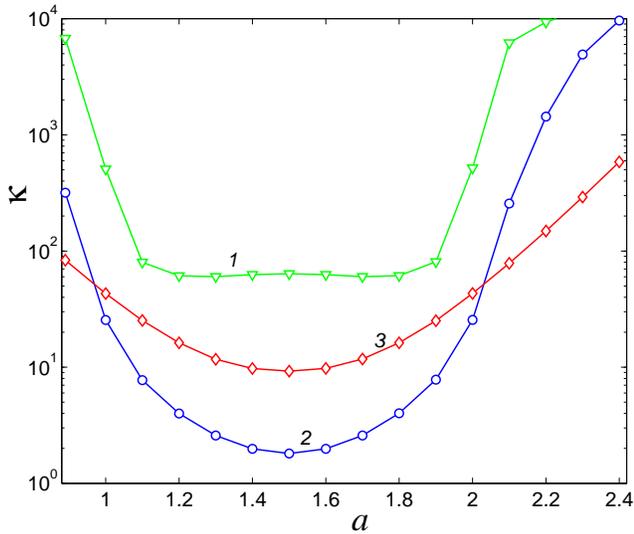}
\caption{
Dependence of the heat conduction coefficient $\kappa$ in the chain with single-well potential
of the nearest-neighbor interaction (\ref{f6}) on the average link length $a$ for temperatures
$T=0.002$, 0.02, 0.2 (curves 1, 2, 3 respectively). Note a logarithmic scale for $\kappa$.
Simulated chain contained in all cases $N_++N+N_-$ particles, $N=640$ of them were free and
$N_\pm=40$ were immersed in the Langevin thermostats with temperatures $T_\pm=(1\pm 0.1)T$.
}
\label{fg12}
\end{figure}

The dependence of the heat conduction coefficient $\kappa$ on the average link length $a$ for
the chain with the single-well non-convex potential of the nearest-neighbor interaction (\ref{f6})
is presented in Fig.~\ref{fg12}. One can observe extremely significant effect of the chain
extension on the heat conduction coefficient. This drastic reduction is observed for values
of the average link length approximately in the
interval $1<a<2$, in accordance with earlier findings on relationship between
a number of possible degenerate (or close energetically) ground states and the reduction
of the heat conductivity.

It is possible to say that these drastic modifications of the heat conduction coefficient occur
due to "latent" double-well nature of the non-convex potential of the nearest-neighbor interaction.
Notably, the most efficient reduction is achieved when the temperature is close to the height
$\epsilon$ of the "latent" potential barrier. The reduction of the heat conductivity by two
orders of magnitude is achieved for $T=0.02$.

It is important to mention that the effective non-convex single-well potential is obtained
in effective models of some quasi-one-dimensional objects. For instance, it is realized in
$\alpha$-spirals of protein macromolecules, double helix of DNA \cite{Savin_pnas,smko13},
as well as in a model of intermetallic NiAl crystalline nanofilms \cite{Dmitriev}.
In these structures the external extension brings a non-homogeneous equilibrium configurations.
Then, one should expect strong effect of external mechanic loads on transport properties
in systems of this sort.

\section{Control in models with convergent heat conduction}
All simple models considered above are commonly believed to have the divergent heat conductivity
in the thermodynamical limit. In this section, we are going to consider a modification of chain
of rotators, which allows external control of the heat conduction coefficient. Simple chain
of rotators is believed to exhibit convergent heat conductivity \cite{giardina,savin2}.
\begin{figure}[tb]
\includegraphics[angle=0, width=1\linewidth]{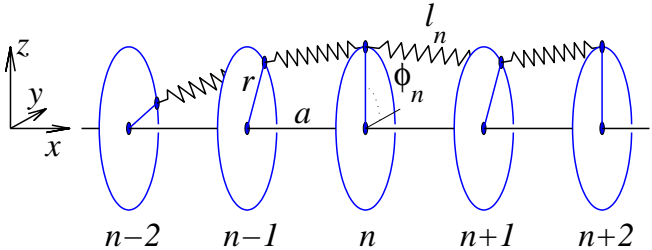}
\caption{Sketch of the modified chain of rotators}
\label{fg13}
\end{figure}

To explain the idea of possible modification, let us consider a mechanical model sketched in Fig.~\ref{fg13}
-- a chain of equal parallel disks of radius $r$ with centers fixed at equal intervals of length $a$
along $x$ axis. The disks can rotate along $x$ axis and the rotation angle of the $n$-th disk is
denoted as $\phi_n$. Neighboring disks are coupled by harmonic springs of equal stiffness and
equilibrium length. Hamiltonian of such model can be simply expressed as follows:
\begin{equation}
H=\sum_{n=1}^N\frac12 I\dot{\phi}_n^2+V(\phi_{n+1}-\phi_n),
\label{f7}
\end{equation}
where $I$ is a moment of inertia of the disk. Potential energy of relative rotation appears due
to deformation of the springs, which couple the neighboring disks, and is expressed as
\begin{eqnarray}
V(\Delta\phi_n)=\frac12K(l_n-L_0)^2\nonumber\\
=\frac12K\{[a^2+2r^2(1-\cos(\Delta\phi_n)]^{1/2}-L_0\}^2,
\label{f8}
\end{eqnarray}
Here $\Delta\phi_n=\phi_{n+1}-\phi_n$ is a relative rotation angle of the neighboring disks,
$K$ and $L_0$ are the stiffness and equilibrium length of the springs respectively, $l_n$ is the
length of the $n$-th spring.
Potential function (\ref{f8}) will be double-well provided that $a<L_0<[a^2+4r^2]^{1/2}$.
The potential minima in this case correspond to the following values of the relative rotation angle:
$$
\Delta\phi_0=\pm 2 \arcsin\{[(L_0^2-a^2)/4r^2]^{1/2}\}.
$$
\begin{figure}[tb]
\includegraphics[angle=0, width=1\linewidth]{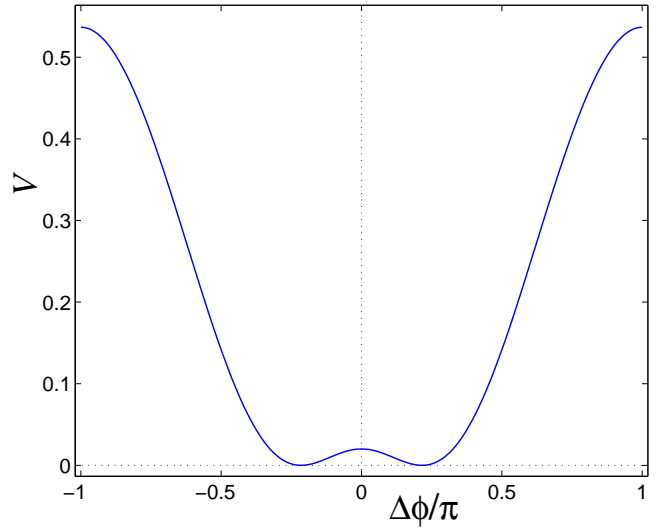}
\caption{
Shape of potential function (\ref{f8}) $V(\Delta\phi)$ for $a=1$, $r=1$, $L_0=1.2$, $K=1$.
Heights of the potential barriers are $\epsilon_0=0.02$, $\epsilon_1=0.5367$.
The potential function achieves the minima for $\Delta\phi=\pm 0.2152\pi$.
}
\label{fg14}
\end{figure}

Characteristic shape of the modified potential function (\ref{f8}) is presented in Fig.~\ref{fg14}.
Without affecting the generality, one can set $I=1$,
$a=1$ and $K=1$. To be specific, we also choose $r=1$. Thus, the potential will have two wells
for the equilibrium length of the spring in the interval $1<L_0<\sqrt{5}$.
The function $V(\Delta\phi)$ is $2\pi$-periodic and has two potential barriers $0<\epsilon_0<\epsilon_1$.
If the equilibrium spring length $L_0$ will only slightly exceed the distance $a$ between the disk centers,
then one will obtain $\epsilon_0\ll\epsilon_1$. For instance, for $L_0=1.2$ one obtains the
potential minima $\Delta\phi_0=\pm0.2152\pi\approx 0.6761$,
and the barriers $\epsilon_0=0.02$, $\epsilon_1=0.5367$.

We expect that, similarly to the simple chain of dipole rotators \cite{giardina,savin2},
the chain with Hamiltonian (\ref{f7}) will have converging heat conduction coefficient.
\begin{figure}[tb]
\includegraphics[angle=0, width=1\linewidth]{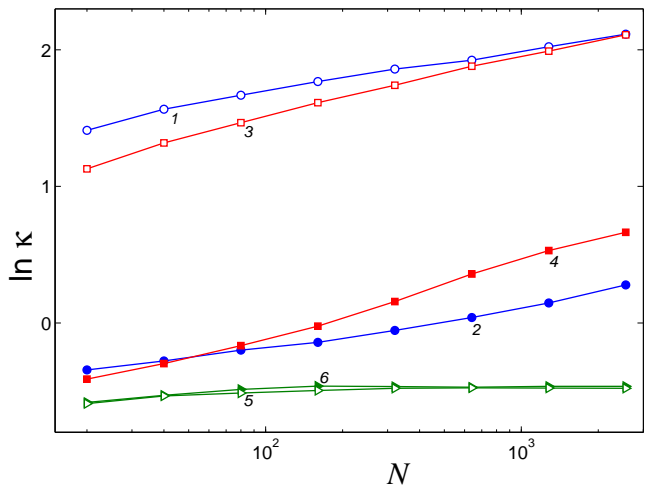}
\caption{
Dependence of the heat conduction coefficient $\kappa$ on the size
of the internal chain fragment $N$ for the chain with periodic potential (\ref{f8})
for the average angle between the disks $\varphi_0=\Delta\phi_0$ and $\varphi_0=0$ respectively
(curves 1 and 2 for temperature $T$=0.01, curves 3 and 4 -- for $T=0.02$,
curves 5 and 6 -- for $T=0.2$).
}
\label{fg15}
\end{figure}

To simulate the heat conductivity, we attach the chain ends with $N_\pm=40$ disks to Langevin
thermostats with temperatures $T_\pm=(1\pm 0.1)T$. Then, we simulate numerically the system of
equation (\ref{f3}) with potential function obtained from (\ref{f8}) under fixed boundary
conditions  $\varphi_0\equiv 0$, $\varphi_M\equiv (N_++N+N_--1)\varphi_0$ and with initial conditions
$$
\{ \varphi_n(0)=(n-1)\varphi_0,~~\dot{\varphi}_n(0)=0\}_{n=1}^{N_++N+N_-},
$$
where $\varphi_0$ -- average value of angle between neighboring disks. In other terms, we keep
constant relative rotation angle between the ends of the chain (in other terms, we apply constant
momentum to the chain) and study the dependence of the heat conduction coefficient on
these external conditions.

Dependence of the heat conduction coefficient $\kappa$ on the length of the chain fragment
between the Langevin thermostats $N$ for the chain with periodic potential (\ref{f8})
is presented in Fig.~\ref{fg15}. As one can see, the value of $N$ for which the heat conductivity
converges, strongly depends on the average temperature $T$.

For average temperature $T=0.2$ comparable to a value of the higher potential barrier $\epsilon_1$,
the convergence is reliably achieved already for $N=320$. We also see that the heat conduction
coefficient almost does not depend on $\varphi_0$. The latter result seems natural, since the
temperature is large enough to allow frequent transitions over the higher barrier; then, the
initial mutual rotation of the disks is relaxed.

If the average temperature will be much lower than the higher potential barrier $T\ll\epsilon_1$,
the relaxation mentioned above will require exponentially large time. In this case, the heat
conduction still converges, but the convergence requires consideration of essentially larger
values of the chain length $N$. For every chain length, the heat conductivity in the chains
with $\varphi_0=0$ is significantly higher that in the chains with $\varphi_0=\Delta\phi_0$.
For the average temperature $T=0.01$ the difference is 6.5 times, and for $T=0.02$ --  5 times.
\begin{figure}[tb]
\includegraphics[angle=0, width=1\linewidth]{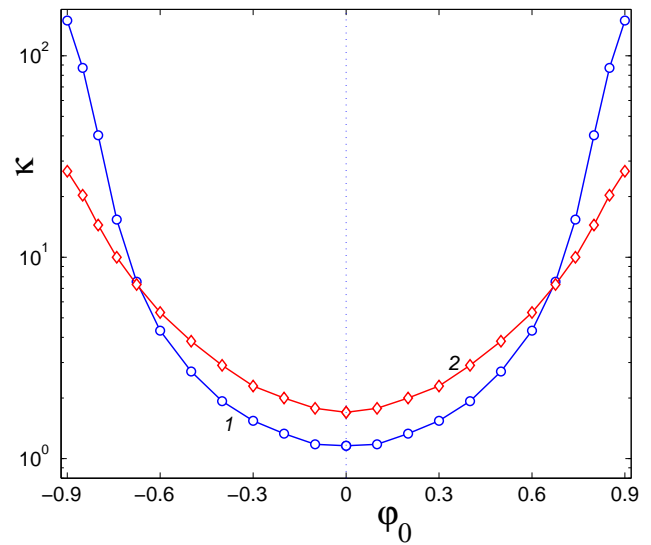}
\caption{
(a) Dependence of the heat conduction coefficient $\kappa$ in the chain with
periodic potential of the nearest-neighbor interaction (\ref{f8}) on the
average interdipole angle $\varphi_0$ for temperatures $T=0.01$, 0.02
(curves 1, 2 respectively).
}
\label{fg16}
\end{figure}

To check this assumption further, let us consider also the dependence of $\kappa$ on the initial
angle between the disks $\varphi_0$ for the chain length $N=1280$. In Fig.~\ref{fg16}, one can
clearly observe that this dependence is strong enough, provided that the temperature is
sufficiently low to prevent the fast relaxation.

Physical reasons of this behavior are very similar to those described in the preceding Sections.
For $|\varphi_0|\le\Delta\phi_0$ the chain has energetically degenerate ground states, where part
of the neighboring disk pairs have relative equilibrium angle $\phi_{n+1}-\phi_n=\Delta\phi_0$,
and the other pairs -- $\phi_{n+1}-\phi_n=-\Delta\phi_0$. A number of possible different ground
states grows as $|\varphi_0|$ tends to zero. That is why the heat conductivity decreases in this limit.
This effect becomes more pronounced as the temperature decreases. Still, for extremely small
temperatures the heat conductivity will cease to depend on $|\varphi_0|$, since even the
smaller potential barriers will become prohibitive.

For exponentially large simulation times the dependence of the heat conductivity on the initial
relative rotation of the disks is expected to disappear due to thermally activated relaxation over
higher potential barriers. In our simulations the times were not large enough to observe this
relaxation for lower temperatures. Still, we believe that this problem may be easily overcome. It seems
sufficient to act with constant external momentum on the right end of the chain, rather than to fix it.
Then, one should investigate a dependence of the heat conduction coefficient on the value of this external momentum.

The idea presented in this Section could be useful for practical design of the systems with
controlled heat conductivity. One can use, for instance, the stretched polymer macromolecules.
Their heat conductivity can be modified by application of the external momentum.

\section*{Concluding remarks}

In this paper we demonstrated that one can efficiently control the transport properties of model
atomic chains by purely mechanical means. In the case of the chain with double-well interparticle
potential, it is enough to change the average interparticle distance in order to modify a number
of possible degenerate ground states and to reduce or to increase the heat conductivity.
The effect is rather pronounced (about five-fold reduction was observed). However, even stronger
effect -- reduction by two orders of magnitude -- was observed in more realistic model with
single-well nonconvex interparticle potential. In this model the heat conductivity is directly
related to applied external strain. Also in this case it seems that the reduction effect is
caused by formation of "effective" double-well potential and variation of a number of possible
states of mechanical equilibrium.

Modification of thermal conductivity in conditions of external mechanical load can be related to
broader field of thermoelasticity. It is well-known that elasticity and plasticity in real
materials can be strongly coupled with thermodynamic phenomena and heat transport. It seems,
however, that this possible coupling has not received proper attention in numerous recent studies
devoted to microscopic foundations of the heat conductivity. Our results demonstrate that these
effects can be rather profound. One can be tempted to say that the extension of the chain
with the DW potential exemplifies the effect of plasticity on the heat transport. The chain
with the single-well potential requires constant external forcing to reduce the heat conductivity;
this phenomenon is primarily elastic. Needless to say, these statements are quite crude and
schematic; still we believe that the subject is of considerable fundamental interest. Besides,
interesting practical implications are straightforward -- one can be interested in simple
mechanical means of control over heat transport in micro- and nanosystems.

\section*{Acknowledgements}

A. V. S. thanks the Joint Supercomputer Center of the Russian Academy of Sciences
for the use of computer facilities.


\end{document}